\documentclass[%
 reprint,
superscriptaddress,
 amsmath,amssymb,
 prapplied,
]{revtex4-2}

\usepackage{graphicx}
\usepackage{dcolumn}
\usepackage{bm}
\usepackage{braket}
\usepackage{hyperref}
\hypersetup{
    colorlinks=true,
    linkcolor=blue,     
    urlcolor=blue,
    citecolor=blue,
}
\usepackage[capitalise]{cleveref}
\usepackage[euler]{textgreek}
\usepackage{xcolor}
\usepackage[version=4]{mhchem}
\usepackage{physics}
\usepackage{multirow}
\usepackage{amsmath}
\usepackage{verbatim}
\usepackage{natbib}
\usepackage{float}
\newcommand{\RNum}[1]{\uppercase\expandafter{\romannumeral #1\relax}}

\begin{document}


\title{Spectroscopy Study on NV Sensors in Diamond-based High-pressure Devices}

\author{Kin On Ho}
 \thanks{These authors contributed equally to this work.}
\affiliation{
Department of Physics, The Chinese University of Hong Kong, Shatin, New Territories, Hong Kong, China
}
\author{Man Yin Leung}
 \thanks{These authors contributed equally to this work.}
\affiliation{
Department of Physics and the IAS Centre for Quantum Technologies, The Hong Kong University of Science and Technology, Clear Water Bay, Kowloon, Hong Kong, China
}
\author{Wenyan Wang}
 \thanks{These authors contributed equally to this work.}
\author{Jianyu Xie}
\author{King Yau Yip}
\affiliation{
Department of Physics, The Chinese University of Hong Kong, Shatin, New Territories, Hong Kong, China
}
\author{Jiahao Wu}
\affiliation{
Department of Physics and the IAS Centre for Quantum Technologies, The Hong Kong University of Science and Technology, Clear Water Bay, Kowloon, Hong Kong, China
}
\author{Swee K. Goh}
\affiliation{
Department of Physics, The Chinese University of Hong Kong, Shatin, New Territories, Hong Kong, China
}
\affiliation{
Shenzhen Research Institute, The Chinese University of Hong Kong, Shatin, New Territories, Hong Kong, China
}
\author{Andrej Denisenko}
\author{J\"{o}rg Wrachtrup}
\affiliation{3rd Institute of Physics, University of Stuttgart and Institute for Quantum Science and Technology (IQST), Pfaffenwaldring 57, D-70569, Stuttgart, Germany}
\author{Sen Yang}
 \email{phsyang@ust.hk}
\affiliation{
Department of Physics, The Chinese University of Hong Kong, Shatin, New Territories, Hong Kong, China
}
\affiliation{
Department of Physics and the IAS Centre for Quantum Technologies, The Hong Kong University of Science and Technology, Clear Water Bay, Kowloon, Hong Kong, China
}

\date{\today}

\begin{abstract}

High-pressure experiments are crucial in modern interdisciplinary research fields such as engineering quantum materials, yet local probing techniques remain restricted due to the tight confinement of the pressure chamber in certain pressure devices. Recently, the negatively charged nitrogen-vacancy (NV) center has emerged as a robust and versatile quantum sensor in pressurized environments. There are two popular ways to implement NV sensing in a diamond anvil cell (DAC), which is a conventional workhorse in the high-pressure community: create implanted NV centers (INVs) at the diamond anvil tip or immerse NV-enriched nano-diamonds (NDs) in the pressure medium. Nonetheless, there are limited studies on comparing the local stress environments experienced by these sensor types as well as their performances as pressure gauges. In this work, by probing the NV energy levels with the optically detected magnetic resonance (ODMR) method, we experimentally reveal a dramatic difference in the partially reconstructed stress tensors of INVs and NDs incorporated in the same DAC. Our measurement results agree with computational simulations, concluding that INVs perceive a more non-hydrostatic environment dominated by a uniaxial stress along the DAC axis. This provides insights on the suitable choice of NV sensors for specific purposes and the stress distribution in a DAC. We further propose some possible methods, such as using NDs and diamond nanopillars, to extend the maximum working pressure of quantum sensing based on ODMR spectroscopy, since the maximum working pressure could be restricted by non-hydrostaticity of the pressure environment. Moreover, we explore more sensing applications of the NV center by studying how pressure modifies different aspects of the NV system. We perform a photoluminescence (PL) study using both INVs and NDs to determine the pressure dependence of the zero-phonon line (ZPL), which helps developing an all-optical pressure sensing protocol with the NV center. We also characterize the spin-lattice relaxation ($T_1$) time of INVs under pressure to lay a foundation for robust pulsed measurements with NV centers in pressurized environments.

\end{abstract}

\maketitle


\section{\label{sec:Intro}Introduction}

Pressure is an important thermodynamic parameter for engineering quantum materials because it allows one to tune material properties without altering the chemical composition, and some long-searched-for quantum phases are expected to emerge under ultra-high pressure, such as room-temperature superconductivity \cite{Drozdov2015Conventional,Maddury2019Evidence,Drozdov2019Superconductivity,Kong2021Superconductivity} and metallic hydrogen \cite{Weir1996Metallization, Loubeyre2020Synchrotron}. High-pressure experiments are, however, non-trivial to perform since one needs robust pressure devices and compatible measurement techniques.

One reliable pressure device is the diamond anvil cell (DAC), which has been widely used in the high-pressure community. The pressure is achieved by mechanically pressing two opposing diamond anvils towards a tightly confined pressure chamber in the middle. The pressure medium filling up the chamber remains hydrostatic below its critical pressure $P_\mathrm{c}$, and it undergoes solidification or glass transition at $P_\mathrm{c}$ where pressure inhomogeneity starts building up. It is crucial to understand whether the medium is hydrostatic during the experiment, since the subsequent data processing and interpretation may be inappropriate if the artifacts from pressure inhomogeneity are not taken into account.

As to the suitable measurement techniques in high-pressure experiments, quantum sensing with negatively charged nitrogen-vacancy (NV$^{-}$) centers has emerged as a strong candidate. We always denote NV$^{-}$ as NV in the rest of this paper. The NV center is a color defect in diamond which consists of a substitutional nitrogen atom, an adjacent atomic vacancy, and an extra electron. Its ground state is an electron spin $S=1$ system and the spin sublevels are responsive to temperature, stress field, magnetic field, electric field, and the surrounding spin bath, making the NV center a versatile sensor for these physical quantities \cite{Acosta2019Color, Barry2020Sensitivity, Schirhagl2014Nitrogen, Jelezko2006Single, Doherty2011The, Ho2021Diamond, Ho2021Recent, Gali2019Ab}. In practice, we measure the electron spin resonance (ESR) spectrum of the NV center using the optically detected magnetic resonance (ODMR) method, which relies on the spin-state-dependent fluorescence rate of the NV center caused by the spin-state-dependent decay route back to the ground state (\cref{fig1}(a)). In ODMR spectroscopy, a green laser is used for initialization and read-out of the NV spin state while a microwave (MW) is used for spin-state manipulation. The ODMR spectrum encodes information about the NV energy structure and hence the environment around the NV center. Due to the superior resolution and sensitivity, quantum sensing with NV centers has become a promising experimental technique.

It has been demonstrated that NV sensing is highly compatible with DACs, and NV centers have outstanding sensing performance even under the demanding conditions inside DACs \cite{Doherty2014Electronic, Steele2017Optically, Yip2019Measuring, Lesik2019Magnetic, Hsieh2019Imaging, Ho2020Probing, Ivady2014Pressure, Toraille2020Combined, Wang2021Ac}. There are mainly two ways to incorporate NV sensors in DACs: (1) create a layer of implanted NV centers (INVs) at a suitable depth inside the diamond anvil tip \cite{Lesik2019Magnetic, Hsieh2019Imaging, Toraille2020Combined}, (2) drop-cast some NV-enriched nano-diamonds (NDs) at the pressure medium interface inside the pressure chamber \cite{Steele2017Optically, Yip2019Measuring, Ho2020Probing}. In general, they are employed to study different kinds of materials under pressure. INVs are commonly used to probe 2D or 3D materials with flat surfaces, while NDs are often applied to examine materials with irregular surfaces. Moreover, INVs and NDs have their own advantages and drawbacks in sensing applications. INVs provide an easy way to detect vector fields because of the known orientation of the bulk diamond crystal in the laboratory frame, yet, the spatial resolution is restricted by the optical diffraction limit and the spatial uniformity of INVs is constrained by imperfections in the implantation procedures; on the other hand, NDs present high spatial resolution controlled by the ND size given the NDs are sparsely distributed and NV centers in the NDs are in close proximity to the sample, yet, the crystal orientations of NDs are random and require individual calibration in the laboratory frame and spin decoherence times of NDs are generally shorter than INVs. Some obvious pros and cons of INVs and NDs are long known, but to the best of our knowledge, no studies have directly compared the pressurized environments perceived by these two types of NV sensors in a single DAC. This incomplete understanding of the pressurized environments at different locations in a DAC may hinder the accurate choice of NV sensors for different experimental purposes.

Another prevailing question from the NV community is the maximum pressure that NV centers can work with as quantum sensors, especially as magnetic field sensors since the probing of local magnetic fields with high spatial resolution is crucial for material research and phase transition studies \cite{Hsieh2019Imaging, Lesik2019Magnetic, Yip2019Measuring}. Ultra-high pressure can bring detrimental effects on quantum sensing with NV centers, including the quenching of ODMR contrasts due to the spin-sublevel mixing in a non-hydrostatic environment. To realize magnetic field sensing in pressurized systems, some previous studies have demonstrated the use of a bias magnetic field to overcome the effects of uniaxial stresses \cite{Hsieh2019Imaging, Lesik2019Magnetic, Yip2019Measuring}. Nonetheless, a strong bias field is required for large uniaxial stresses. This may impose technical difficulties on the experimental setup, and a strong bias field may undesirably change the properties of the material under investigation. Thus, it is of interest to explore other complementary solutions for extending the working pressure of NV sensing.

In this work, we first incorporate both INVs and NDs in the same DAC and analyze the difference in effective pressure transmissions from the hydrostatic pressure medium to these two types of sensors. We partially reconstruct the local stress tensors perceived by INVs and NDs using information from the respective ODMR spectra, and we also perform finite-element simulations to cross-check our experimental findings. These analyses serve to calibrate the local pressurized environments of the two sensor types, to compare their performances as hydrostatic pressure gauges, and to determine their optimal working ranges. By comparing the pressure conditions of the two sensor types, we demonstrate how non-hydrostaticity restricts the maximum working pressure of NV sensing, and we further propose some possible solutions to conquer the non-hydrostaticity. Besides, thoroughly characterizing the stress responses of NV sensors may pave the way for simultaneous detection of multiple physical parameters via ODMR spectroscopy, like pressure and temperature or pressure and magnetic field.

Next, we employ our ODMR-calibrated NV sensors to investigate from different perspectives the pressure-tuned energy structure of the NV center. We measure the photoluminescence (PL) spectra of both INVs and NDs to study the pressure dependence of the zero-phonon line (ZPL), which represents the energy spacing between electronic orbitals of the NV center. We also measure the spin-lattice relaxation ($T_1$) time of INVs against pressure to probe the electron-phonon coupling in the solid-state defect system. Combining various spectroscopic techniques ranging from continuous-wave (cw) to pulsed measurements and from ESR to PL measurements, we hereby provide a multi-dimensional understanding of the NV quantum system under high pressure, which helps fostering more accurate and distinct applications of NV sensing in extreme conditions. Such applications include an all-optical pressure sensing protocol based on PL spectroscopy and robust implementation of pulse sequences at high pressure.

\section{\label{sec:Theory} Theoretical Background}

In a single crystalline diamond with an ensemble of NV centers, there are four possible spatial orientations for the NV centers. We thus have five relevant reference frames: the crystal frame $(X,Y,Z)$ and the principal axis frames for the four NV orientations $(x,y,z)^k$, $k\in\{$NV1, NV2, NV3, NV4$\}$. The four NV frames can be related by simple rotation transformations due to the symmetry of the diamond crystal. In this work, we follow Barfuss \textit{et al.}'s conventions of the five frames and the coordinate transformations between them \cite{Barfuss2019Spin}, and we always take compressive stresses to be positive.

The NV center is a robust stress sensor due to the spin-stress coupling effect \cite{Barson2017Nanomechanical, Udvarhelyi2018Spin, Broadway2019Microscopic, Barfuss2019Spin, Hsieh2019Imaging, Ho2020Probing, Ho2021In, Marshall2022High, Gali2019Ab, Doherty2012Theory, Falk2014Electrically, Maze2011Properties, Doherty2011The, Doherty2014Temperature, Ivady2014Pressure, Kobayashi1993High, Deng2014New, Udvarhelyi2018Ab, Kehayias2019imaging}. Under a stress field affecting the spin-spin interaction, the ground-state Hamiltonian for each NV orientation in its principal axis frame can be written as \cite{Barfuss2019Spin, Barson2017Nanomechanical, Udvarhelyi2018Spin}
    \begin{align}
        H^{k} =\;& (D_{0} + M_{z}^{k})S_{z}^{2} + M_{x}^{k}(S_{y}^{2}-S_{x}^{2})\nonumber\\
        &+ M_{y}^{k} \{S_x,S_y\} + N_{x}^{k} \{S_x,S_z\} + N_{y}^{k} \{S_y, S_z\},
    \label{Hk}
    \end{align}
where $S$ is the spin-1 operator, $D_{0}=2870$ MHz in ambient conditions, and in the hybrid representation, the NV-frame quantities $M_{x,y,z}^k$ and $N_{x,y}^k$ can be expressed in terms of the components $\sigma_{IJ}$ of the crystal-frame stress tensor $\bm{\sigma}$. For NV1 along [111] crystal direction,
    \begin{align}
        M_{z}^{\mathrm{NV1}} =\;& a_{1} (\sigma_{XX}+\sigma_{YY}+\sigma_{ZZ})\nonumber\\
        &+ 2a_{2}(\sigma_{YZ}+\sigma_{XZ}+\sigma_{XY}),\label{NV1Mz}\\
        M_{x}^{\mathrm{NV1}} =\;& b(2\sigma_{ZZ}-\sigma_{XX}-\sigma_{YY})\nonumber\\
        &+c(2\sigma_{XY}-\sigma_{YZ}-\sigma_{XZ}),\label{NV1Mx}\\
        M_{y}^{\mathrm{NV1}} =\;& \sqrt{3}b(\sigma_{XX}-\sigma_{YY})+\sqrt{3}c(\sigma_{YZ}-\sigma_{XZ}),\label{NV1My}\\
        N_{x}^{\mathrm{NV1}} =\;& d(2\sigma_{ZZ}-\sigma_{XX}-\sigma_{YY})\nonumber\\
        &+ e(2\sigma_{XY}-\sigma_{YZ}-\sigma_{XZ}),\label{NV1Nx}\\
        N_{y}^{\mathrm{NV1}} =\;& \sqrt{3}d(\sigma_{XX}-\sigma_{YY})+\sqrt{3}e(\sigma_{YZ}-\sigma_{XZ}),
    \label{NV1Ny}
    \end{align}
where $a_1$, $a_2$, $b$, $c$, $d$, and $e$ are the spin-stress coupling constants in the hybrid representation. To obtain the above expressions for the other three NV orientations, we need to replace $\sigma_{IJ}$ by $(\bm{K}^l\cdot\bm{\sigma}\cdot(\bm{K}^l)^\mathrm{T})_{IJ}$ in \cref{NV1Mz,NV1Mx,NV1My,NV1Nx,NV1Ny}, where $\bm{K}^l$ are the coordinate transformations from NV1 to $l\in\{$NV2, NV3, NV4$\}$ as defined in Ref. \cite{Barfuss2019Spin}. The resulting expressions for NV2-4 are different from \cref{NV1Mz,NV1Mx,NV1My,NV1Nx,NV1Ny} only by sign flips in some of the off-diagonal tensor components. See Supplementary Materials for details.

\begin{figure}[t]
\includegraphics[width=8.6cm]{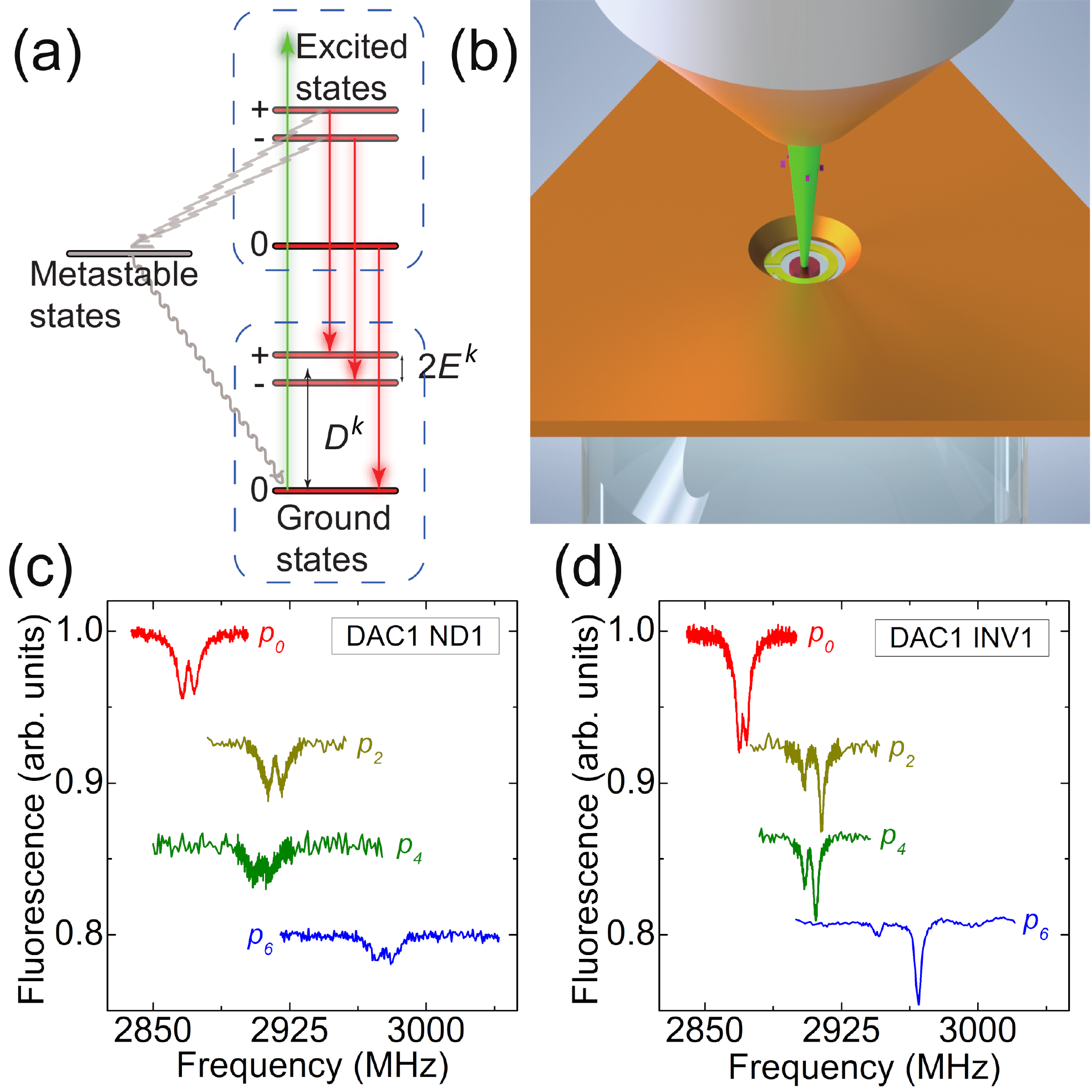}
\caption{(a) A simplified energy level diagram of the NV center showing the spin-state-dependent transitions. (b) An illustration of the DAC configuration in our experiment. The MW antenna is fabricated on the implanted diamond anvil culet, while some 140-nm NDs are drop-casted on the other un-implanted anvil culet. (c, d) ODMR responses of a 140-nm ND (labeled as ND1) and a patch of INVs (labeled as INV1) to the change in pressure of DAC1, respectively. The decrease in their ODMR contrasts is due to the stress-induced mixing of NV spin states and the degradation of the MW structure, where the latter factor takes a heavier toll on ND1.}
\label{fig1}
\end{figure}

Experiments have found that $a_{1} = 0.486\pm0.0002$, $a_{2} = -0.37\pm0.002$, $b = -0.147\pm0.0002$, $c = 0.342\pm0.0007$ MHz/kbar \cite{Barson2017Nanomechanical, Hsieh2019Imaging}, agreeing well with the theoretical values from a density functional theory (DFT) study \cite{Udvarhelyi2018Spin}. This DFT study also reports $d=0.012(1)$ and $e=-0.066(1)$ MHz/kbar. Since $d$ and $e$ are an order of magnitude smaller than the rest of the coupling constants, we can neglect the $N_x^k$ and $N_y^k$ terms in \cref{Hk} for our first-order discussion here, and the three eigenvalues of the Hamiltonian $H^k$ can thus be analytically solved as follows,
    \begin{equation}
        f_0^k = 0,\; f_{\pm}^k = D_0 + M_z^k \pm \sqrt{(M_x^k)^2 + (M_y^k)^2}.
    \end{equation}
Hence, $f_{\pm}^k$ are the two resonance frequencies detectable by ODMR spectroscopy, with their center and splitting being $D^k$ and $2E^k$ respectively (\cref{fig1}(a)).

In the regime of small shear stresses, the four NV orientations have close $f_{+}$'s and close $f_{-}$'s, leading to two overall resonances in the ODMR spectrum of the whole NV ensemble. We further assume equal population for the four NV orientations in the diamond crystal, such that the two overall ODMR resonances should be averages of $f_{+}^k$ and $f_{-}^k$ over $k\in\{$NV1, NV2, NV3, NV4$\}$, with their center $D$ and splitting $2E$ written respectively as
    \begin{align}
        D &= D_0 + \frac{1}{4}\sum_k M_z^k,\label{D}\\
        E &= \frac{1}{4}\sum_k \sqrt{(M_x^k)^2 + (M_y^k)^2}.\label{E}
    \end{align}
These expressions reveal that $D$ scales with pressure, while $E$ results from the imbalance between uniaxial stresses along the three orthogonal directions and the presence of shear stresses, or in other words $E$ is an indicator of hydrostaticity. When we compress the diamond crystal, both $D$ and $E$ will increase in general, i.e. the ODMR resonances will shift to the right and split further apart. 

With \cref{D,E} in hand, we can employ ODMR spectroscopy to partially reconstruct the crystal-frame stress tensor $\bm{\sigma}$ perceived by the NV ensemble. This theory section is applicable for both INVs and NDs, and to have meaningful interpretations of the reconstructed crystal-frame stress tensors, we must also understand how the INV and ND crystal frames are related to the laboratory frame, which we will discuss in \cref{sec:Exp}. 

\section{\label{sec:Exp}Experimental setup}

\cref{fig1}(b) illustrates our customized DAC design where both INVs and NDs are incorporated. We utilize (100)-oriented diamond anvils, and the layer of INVs is located at the culet of one of the anvils. This implanted anvil is prepared by 9.8 keV \ce{^15N} ion implantation at a dose of $\sim$10$^{12}$ N/cm\textsuperscript{2} and subsequent annealing at 950\textsuperscript{o}C in a high vacuum ($P<10^{-6}$ mbar) for 2 hours. The resulting implantation area has a diameter of 200 \textmu m and is at a depth of $\sim$10 nm below the culet surface that has a surface roughness of $\sim$1.5 nm. On the other hand, some 140-nm NDs with nitrogen concentration of 3 ppm are sparsely drop-casted on the culet surface of the other un-implanted diamond anvil. To perform ODMR spectroscopy with these two types of NV sensors, a 150-\textmu m-radius Omega-shaped gold microstructure is fabricated on the implanted anvil for MW transmission \cite{Xie2021Fragile}. As to the pressure chamber in our design, a 300-\textmu m-diameter hole is drilled in the middle of a beryllium-copper gasket and the hole is completely filled with a 4:1 methanol:ethanol mixture as the pressure medium. At room temperature, this particular medium remains hydrostatic up to $\sim$100 kbar \cite{Piermarini1973Hydrostatic, Angel2007Effective, Tateiwa2009Evaluations, Klotz2009Hydrostatic, Ho2022Probing} which fully covers our experimental pressure range, enabling us to compare the local pressurized environments of INVs and NDs given the medium is in an excellent hydrostatic condition. Another reason for choosing this medium is that most of the common phase transitions tuned by pressure in condensed matter physics lie within 100 kbar. Therefore, it is of technical significance to study the stress distribution in a DAC, which is a popular pressure device in material research, using a medium with the hydrostatic limit up to 100 kbar.

We have prepared two DACs based on the above-described design, where all the cell configurations are the same except for the thickness of the pre-indented gasket (150 \textmu m in one DAC and 200 \textmu m in the other). We will denote these two DACs as DAC1 and DAC2 respectively hereafter. In our experiments with the DACs, a home-built confocal microscope with a 520-nm laser diode and a long-working-distance objective (50X Mitutoyo Plan Apo SL) is used to optically address the NV sensors, and the local pressure is calibrated by $\partial D/\partial P = 1.49$ MHz/kbar \cite{Ho2020Probing} and the $D$ value at ambient pressure measured by the corresponding NV sensor (the ambient $D$ values have only tiny deviations from $D_0=2870$ MHz).

Since the implanted anvil is (100)-oriented, it is natural to define the INV crystal frame ($X,Y,Z$) with the $X$ axis along the DAC axis. On the other hand, it is not that trivial to determine how the crystal frames of individual NDs are oriented with respect to the laboratory frame. We need to first apply a static magnetic field along the DAC axis and measure the ODMR spectrum of the target ND. Then by studying the Zeeman splittings in the spectrum, we can obtain the projection angles of the DAC axis onto the four NV orientations \cite{Doherty2014Measuring}. The unit direction of the DAC axis in the ND crystal frame can thus be computed by solving an effective problem of the intersection of three cones (see Supplementary Materials for details). The subsequent stress analysis should not depend on exactly how we assign the four angles to the four NV orientations (NV1-4) under our assumption of the equal population for the four orientations, and we will explicitly check that this is the case in \cref{sec:tensor}.

For PL measurements, we use the 520-nm laser diode to excite NV electrons from the electronic ground state to the phonon band above the electronic excited state. The NV electrons would decay to the zero-phonon mode of the excited state via emitting infrared (IR) radiation, then to the phonon band of the ground state via emitting red PL, and finally to the zero-phonon mode of the ground state via emitting IR radiation \cite{Kehayias2013Infrared}. The ZPL in the resulting PL spectrum is produced by NV electrons that decay from the zero-phonon mode of the excited state directly back to the zero-phonon mode of the ground state. The PL spectra of INVs and NDs are collected using a commercial spectrometer (Princeton Instrument IsoPlane SCT-320) with a 550-nm long-pass filter in front. To obtain a PL spectrum solely originating from the NV centers in a targeted sensor, we subtract the PL spectrum measured under an applied MW field at one of the ODMR resonance frequencies from the spectrum without any exerted MW. This method makes use of the spin-state dependence of the NV fluorescence. To enhance the data quality, we choose to drive whichever one of the two ODMR resonances with higher contrast.

\section{\label{sec:Compare} Comparisons of local pressurized environments}

DAC1 (DAC2) is pressurized in an ascending pressure sequence from the ambient pressure $p_0$ up to $p_6$ ($p_5$), except that $p_4$ ($p_5$) is a reduced pressure point. Throughout the experiment with DAC1 (DAC2), we have tracked three (five) 140-nm NDs and six (six) 500-nm patches of INVs. Note that our confocal microscope has a lateral resolution of $\sim$500 nm, and we will number the tracked sensors in DAC1 and DAC2 with Arabic numerals and in alphabetical order respectively, e.g. ND1, INV1, NDa, INVa. In general, the difference between the local pressurized environments of NDs and INVs becomes more significant as we increase the DAC pressure.

Using data from DAC1 as examples, we present in \cref{fig1}(c) and (d) how the raw ODMR spectra of ND1 and INV1 change with the DAC1 pressure respectively. Their spectral changes can be compared in terms of the center $D$ and splitting $2E$ of the ODMR resonances. At $p_0$, ND1 and INV1 agree well on $D$ but ND1 has a larger $E$ than INV1, indicating a larger intrinsic lattice distortion in ND1. When DAC1 is pressurized to $p_2$, ND1 shows a greater rightward shift in $D$ while INV1 exhibits a more noticeable increase in $E$, and such differences in their spectral responses become more pronounced at $p_6$. These reveal that when we press the diamond anvils towards each other, ND1 experiences stronger local pressure from a more hydrostatic environment at the pressure medium interface, while INV1 is subjected to weaker local pressure from a more directional stress environment inside the anvil culet. The stress anisotropy around INV1 may have produced a symmetry breaking between the two ground-state sublevel transitions, as seen from the increasingly unequal contrasts of the two ODMR resonances at $p_2$ and $p_6$ in \cref{fig1}(d). On the other hand, both ND1 and INV1 show decreases in $D$ and $E$ at the reduced pressure point $p_4$, reflecting the expected stress relaxation when we loosen the diamond anvils. Note that the decline in ODMR contrasts for ND1 and INV1 is due to the stress-induced mixing of NV spin states and the degradation of MW structure, where the latter factor takes a heavier toll on ND1. Apart from the artifact of the deteriorated MW structure, all the mentioned main features in the ODMR responses of the two NV sensor types can be reproduced in the independent experiment with DAC2 (see Supplementary Materials).

\begin{figure}[t]
\includegraphics[width=8.6cm]{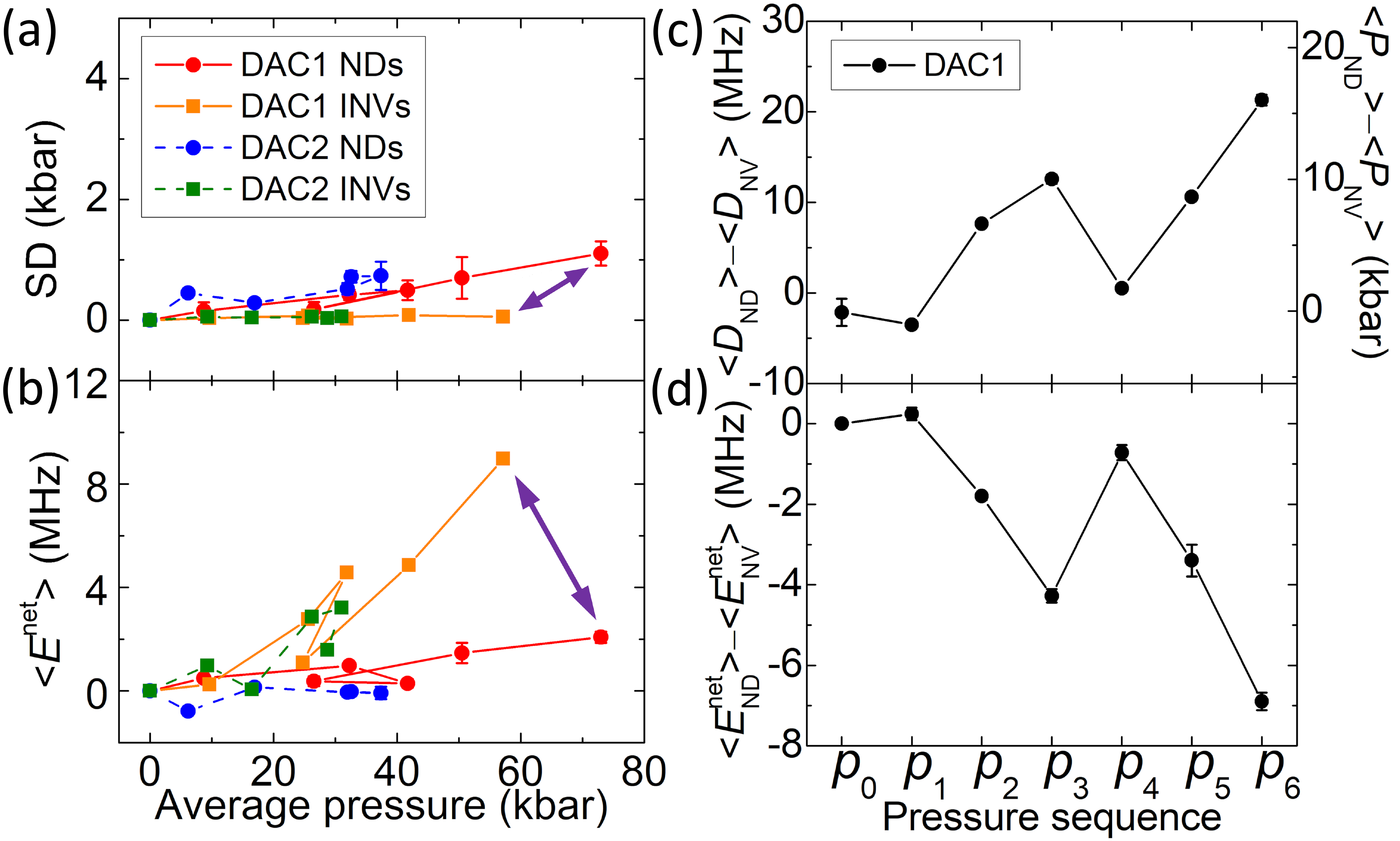}
\caption{(a, b) Plots of SD of pressure and average $E^{\mathrm{net}}$ against average pressure for three (five) NDs and six (six) patches of INVs in DAC1 (DAC2), where $E^{\mathrm{net}}$ is the net change in $E$ with respect to the ambient value. The two DACs reveal very similar data trends. The NDs only show a tiny increase in SD while the INVs have no observable SD at all, signifying the pressure homogeneity at the pressure medium interface and $\sim$10 nm deep in the culet. Moreover, the INVs reveal a much greater increase in average $E^{\mathrm{net}}$, implying a more hydrostatic environment around the NDs and a more anisotropic environment around the INVs. Here, the markers are joined in a way to indicate the pressure sequences in the experiments, while the purple arrows drawn are to emphasize the significantly different behaviors of NDs and INVs at the highest pressure point achieved. (c, d) The differences in average $D$, average $P$, and average $E^{\mathrm{net}}$ between the three NDs and the six patches of INVs in DAC1 along the pressure sequence. In general, the NDs experience much stronger hydrostatic pressure compared with the INVs. For all subfigures, one of the three NDs in DAC1 is replaced by another ND for the statistics at $p_{4}$ and $p_{6}$, due to the occasionally weakened fluorescence of those NDs.}
\label{fig2}
\end{figure}

To go beyond describing the raw spectra, we perform statistical comparisons of the local environments perceived by NDs and INVs. In \cref{fig2}(a) and (b), we plot the standard deviation (SD) of pressure and average $E^{\mathrm{net}}$ against average pressure for the tracked NDs and the tracked patches of INVs in DAC1 and DAC2, where $E^{\mathrm{net}}$ is the measured $E$ offset by the ambient value. It is evident that the two DACs give rise to very similar results. First, the NDs only show a tiny increase in the SD of pressure while the patches of INVs have no observable SD at all. This suggests we have highly homogeneous pressure at both the medium interface and $\sim$10 nm deep in the culet, and the small SD from the NDs also hints at an excellent hydrostatic condition of the pressure medium (4:1 methanol:ethanol mixture) within the pressure range under investigation \cite{Ho2020Probing, Ho2022Probing}. Second, the average pressure detected by the NDs becomes increasingly greater than that detected by the patches of INVs, and the patches of INVs have a much more remarkable increase in the average $E^{\mathrm{net}}$ compared with the NDs. These statistics verify our previous inference that a more hydrostatic environment exists at the medium interface to produce stronger local pressure, while a more anisotropic environment exists inside the anvil culet to give weaker local pressure. Third, at the reduced pressure points, the NDs and the patches of INVs show much smaller differences in the average pressure, SD of pressure, and average $E^{\mathrm{net}}$. This implies relaxation of the DAC may tend to ``unify'' the pressurized environments at the medium interface and inside the culet. Note that for DAC2, the data of NDs at $p_1$ may be affected by the instability of the pressure medium due to insufficient buffer time between the pressurization of the DAC and measurements. As a more direct comparison to supplement the above discussions, we plot in \cref{fig2}(c) and (d) the quantitative differences in average $D$, average pressure $P$, and average $E^{\mathrm{net}}$ between the NDs and the patches of INVs in DAC1 along the pressure sequence.

\section{\label{sec:tensor} Quantitative stress tensor analysis}

In this section, we will consider net effects of the DAC pressure on the stress tensors of the two NV sensor types.

First, we assume the stress tensors induced by the DAC pressure to be quasi-hydrostatic, i.e. a hydrostatic pressure $P$ plus a first-order correction from a uniaxial stress of magnitude $\lambda P$ along the DAC axis. This assumption is intuitive since any non-hydrostaticity in the DAC is likely to arise from the symmetry breaking due to the external force applied along the DAC axis. Under this assumption, the crystal-frame stress tensor of an ND can be written as
    \begin{gather}
        \sigma^{\mathrm{ND}}
        =
        \begin{pmatrix}
         P & 0 & 0 \\
         0 & P & 0 \\
         0 & 0 & P
        \end{pmatrix}
        +
        \bm{U}_{\bm{n}}^\mathrm{T}
        \begin{pmatrix}
         \lambda P & 0 & 0 \\
         0 & 0 & 0 \\
         0 & 0 & 0
        \end{pmatrix}
        \bm{U}_{\bm{n}},
    \label{sigmaND}
    \end{gather}

\begin{figure}[H]
\includegraphics[width=8.6cm]{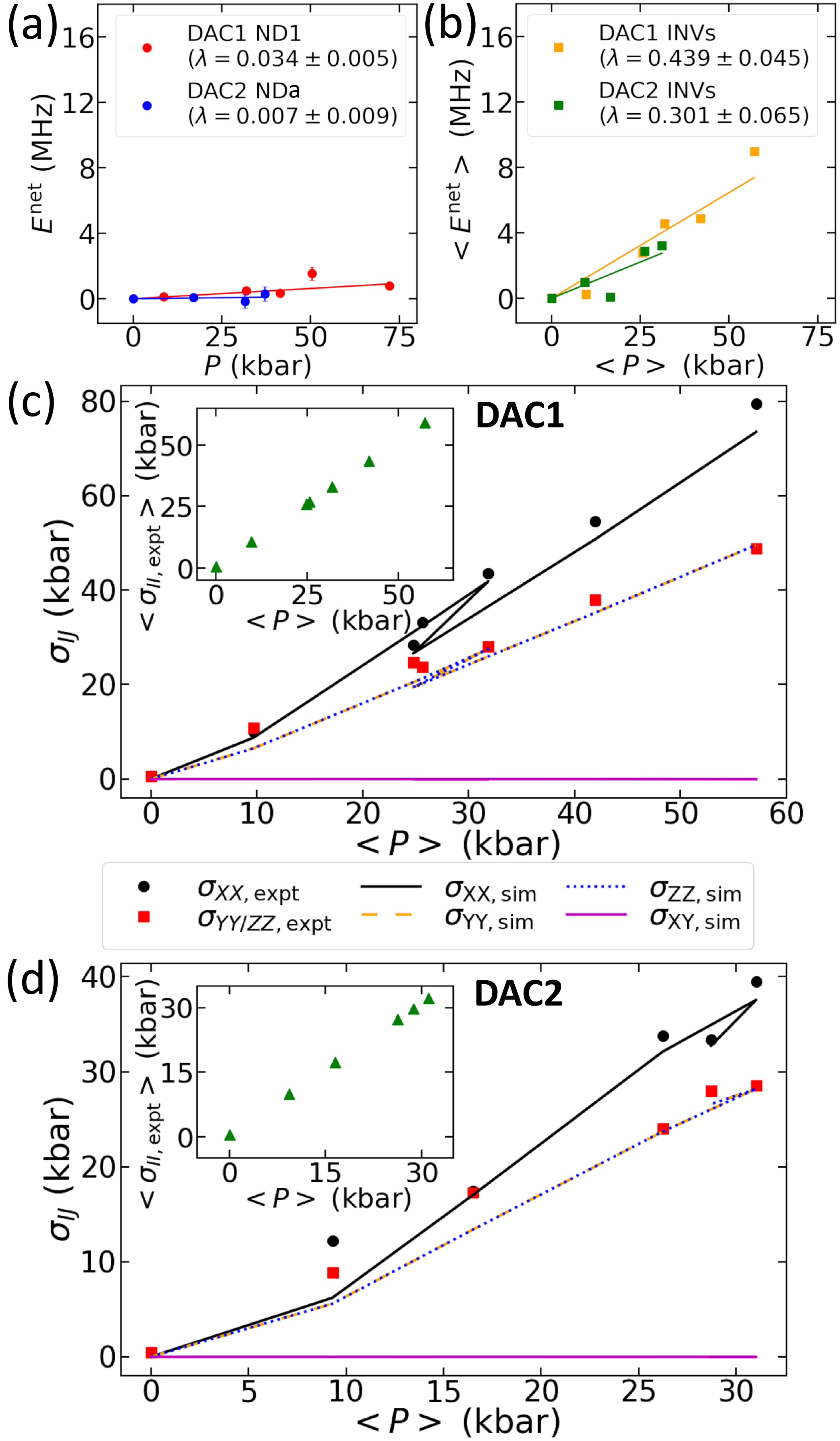}
\caption{(a) Linear fittings of $E^{\mathrm{net}}$ versus $P$ for ND1 in DAC1 and NDa in DAC2 by putting \cref{sigmaND} into \cref{E}. (b) Linear fittings of average $E^{\mathrm{net}}$ versus average $P$ for the six patches of INVs in DAC1 and DAC2 by substituting \cref{sigmaINV} into \cref{E}. The fitted values of $\lambda$ for the INVs are at least an order of magnitude greater than that of the corresponding ND in the same DAC, suggesting a more directional stress environment around the INVs with a dominant uniaxial contribution along the DAC axis. Note that the reduced pressure points are not included in (a) and (b) for concentrating on the effects of pressurizing our DACs, and $p_1$ is omitted in the fitting for NDa in DAC2 owing to the instability of the pressure medium during measurements of NDs at this pressure point. (c, d) INV crystal-frame stress tensor components derived from experimental data (presented with markers) and obtained from computational simulations (presented with lines) against average $P$ measured by the six patches of INVs, for DAC1 and DAC2 respectively. The lines are plotted in a way to signify the pressure sequences in the experiments. Both the experiment and simulation results reveal the gradual dominance of the stress component along the DAC axis. The insets illustrate that the average of our derived diagonal stress components indeed gives the average measured pressure. (c) and (d) share the same legend.}
\label{fig3}
\end{figure}

\noindent
where $\bm{U}_{\bm{n}}$ is the coordinate transformation from the ND crystal frame to an auxiliary frame with its basis vector $\bm{e}_X^{\prime}$ along the direction $\bm{n}$ of the DAC axis observed in the ND crystal frame (see Supplementary Materials for the determination of $\bm{U}_{\bm{n}}$). As stated in \cref{sec:Exp}, we can derive the projection angles of the DAC axis onto the four NV orientations from the Zeeman splittings of the ND under an external magnetic field along the DAC axis. The angles found for ND1 in DAC1 and NDa in DAC2 are $\{16.0^\circ, 51.5^\circ, 79.3^\circ, 79.3^\circ\}$ and $\{22.5^\circ, 59.1^\circ, 59.1^\circ, 85.1^\circ\}$ respectively. For each of these two NDs, there are 12 permutations of assigning these four angles to NV1-4, hence leading to 12 cases of $\bm{n}$ solved from the respective three-cone problems. Note that there would be 24 cases if we have four distinct projection angles instead. As we expected in \cref{sec:Exp}, all the 12 $\bm{n}$ from each ND give rise to the same linear equation when we substitute their corresponding forms of \cref{sigmaND} into \cref{E} ($E(P)=0.366\lambda P$ for ND1, $E(P)=0.340\lambda P$ for NDa). These linear equations are then used to fit the curves of $E^{\mathrm{net}}$ versus $P$ measured by the two NDs (see \cref{fig3}(a)), and the fitted values of $\lambda$ are $0.034\pm0.005$ and $0.007\pm0.009$ for ND1 and NDa respectively. Note that the reduced pressure points are not included in the linear curve fittings since pressurization and relaxation of a DAC are not simply the reverse of one another, as described in Ref. \cite{Ho2020Probing} and \cref{sec:Compare}. Moreover, $p_1$ is omitted in the fitting for NDa in DAC2 owing to the instability of the pressure medium during measurements of NDs at this pressure point.

On the other hand, the INV crystal-frame stress tensor can be expressed as
    \begin{gather}
        \sigma^{\mathrm{INV}}
        =
        \begin{pmatrix}
         P & 0 & 0 \\
         0 & P & 0 \\
         0 & 0 & P
        \end{pmatrix}
        +
        \begin{pmatrix}
         \lambda P & 0 & 0 \\
         0 & 0 & 0 \\
         0 & 0 & 0
        \end{pmatrix},
    \label{sigmaINV}
    \end{gather}
which takes a simpler form since the $X$ axis of the INV crystal frame is precisely defined along the DAC axis as stated in \cref{sec:Exp}. Similarly to the NDs, we substitute \cref{sigmaINV} into \cref{E} and use the resulting linear equation to fit the curves of average $E^{\mathrm{net}}$ versus average $P$ measured by the six patches of INVs in DAC1 and DAC2 (see \cref{fig3}(b)). The fitted values of $\lambda$ are $0.439 \pm 0.045$ and $0.301 \pm 0.065$ respectively, which are at least an order of magnitude greater than that of the corresponding ND in the same DAC. We therefore speculate that the INVs perceive a much more directional stress environment with a dominant uniaxial contribution along the DAC axis.

Next, we conduct a deeper study of the INV crystal-frame stress tensor originating from the DAC pressure, by writing down a more general tensor form,
    \begin{gather}
        \sigma^{\mathrm{INV}}
        =
        \begin{pmatrix}
         \sigma_{XX} & 0 & 0 \\
         0 & \sigma_{YY} & 0 \\
         0 & 0 & \sigma_{ZZ}
        \end{pmatrix},
    \label{sigmaINV2}
    \end{gather}
where we assume negligible shear stresses and a cylindrical symmetry about the DAC axis (i.e. $\sigma_{YY}=\sigma_{ZZ}$). By substituting \cref{sigmaINV2} into \cref{D,E}, we can use the average $D$ and average $E^{\mathrm{net}}$ measured by the six patches of INVs in each DAC to determine the $\sigma_{XX}$ and $\sigma_{YY/ZZ}$ perceived by INVs at each pressure point. The experimentally derived results are presented using markers in \cref{fig3}(c) and (d) for DAC1 and DAC2 respectively, where the loading stress $\sigma_{XX,\mathrm{expt}}$ gradually becomes greater than the lateral stress $\sigma_{YY/ZZ, \mathrm{expt}}$ in both DACs. Quantitatively, the ratio of $\sigma_{XX,\mathrm{expt}}$ to $\sigma_{YY/ZZ, \mathrm{expt}}$ increases from 1 at $p_0$ to 1.62 (1.38) at the highest pressure point for DAC1 (DAC2). This demonstrates the accumulation of non-hydrostaticity in the diamond anvil culet due to the gradual dominance of the stress component along the DAC axis, in accord with our previous claims. To showcase the validity of our results, we further illustrate in the insets of \cref{fig3}(c) and (d) that the average of our derived diagonal stress components, $<\sigma_{II, \mathrm{expt}}>$, indeed gives the average pressure measured by the six patches of INVs.

\begin{figure}[t]
\includegraphics[width=8.6cm]{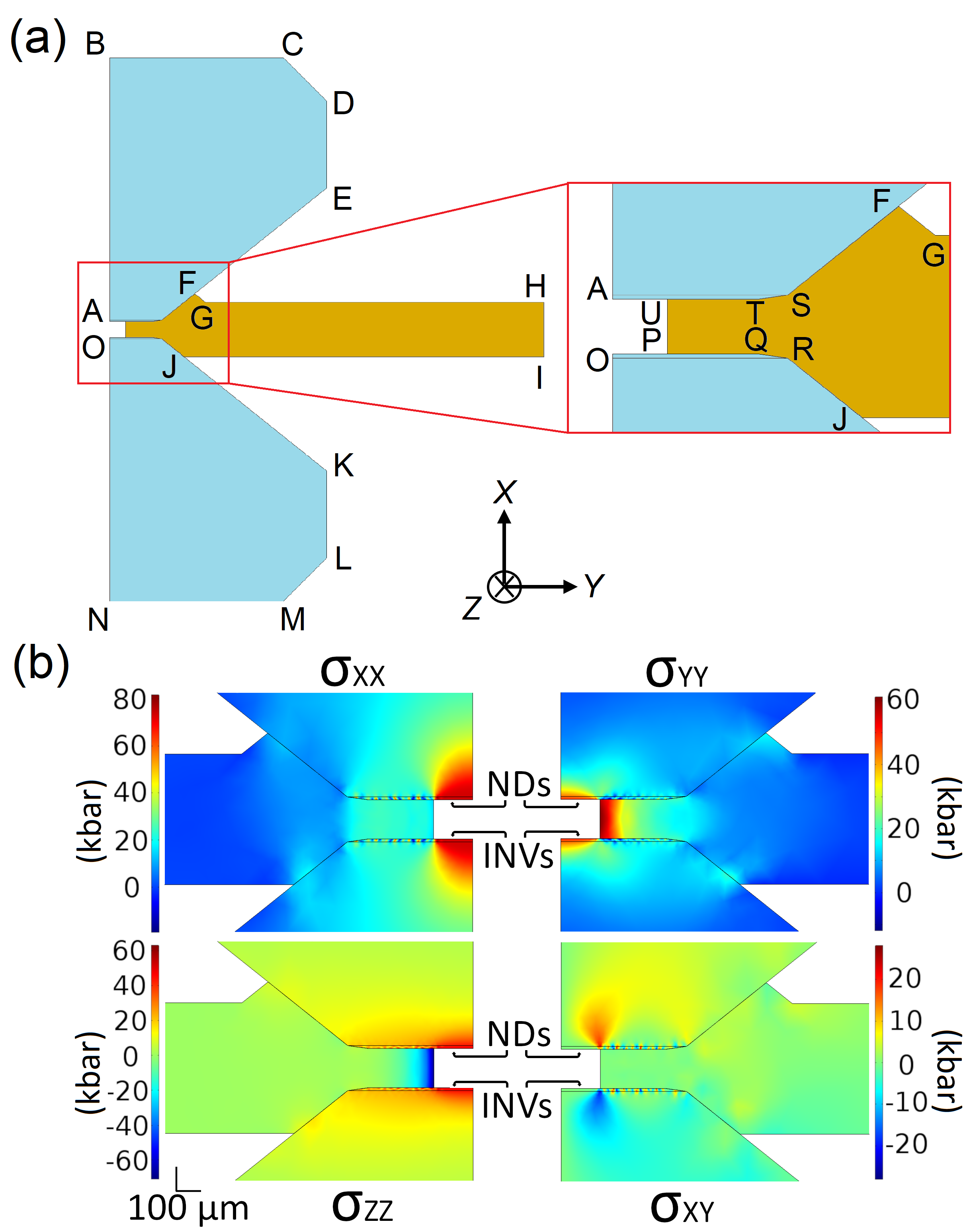}
\caption{(a) The 2D axisymmetric model used in our computational simulations, with the blue regions representing the (100)-oriented diamond anvils (the bottom one is the implanted anvil) and the orange-brown region representing the beryllium-copper gasket. A zoom-in on the beveled diamond culets is provided, and the orientation of INV crystal frame with respect to the DAC model is specified. All the boundaries are properly named for discussing the boundary conditions in our simulations. (b) Colour maps visualizing the spatial distributions of the loading stress $\sigma_{XX}$, the lateral stresses $\sigma_{YY}, \sigma_{ZZ}$, and the shear stress $\sigma_{XY}$ from the simulation at the highest pressure point $p_6$ of DAC1. Compressive stresses are taken to be positive. The distributions of $\sigma_{XX}$, $\sigma_{YY}$, and $\sigma_{ZZ}$ are symmetric in the two anvils but it is not the case for $\sigma_{XY}$. Furthermore, $\sigma_{XY}$ is considerably smaller than the diagonal stress components. }
\label{fig4}
\end{figure}

To cross-check the above INV crystal-frame stress tensors derived from our experimental data, we perform simulations using a finite-element analysis software. We employ the solid mechanics module in the software to study the steady-state problem at each pressure point for our DACs. \cref{fig4}(a) shows the 2D axisymmetric model used in our simulations, which consists of the two (100)-oriented diamond anvils and the beryllium-copper gasket (the bottom anvil is the implanted one). The anvil geometry follows the standard design of a Type IIas diamond anvil from the manufacturer Megabar-Tech, with the culet diameter being 800 \textmu m. The $X$-axis of the INV crystal frame is defined along the DAC axis as usual, and we impose the following boundary conditions in the simulation at each pressure point for our DACs (refer to \cref{fig4}(a) for the naming of boundaries):

(1) Boundary loads: the base of the un-implanted anvil (boundary BC) is loaded with our externally applied force, while the pressure medium interfaces (boundaries AU, UP, and PO) are loaded with the average pressure measured by our NDs.

(2) Displacement constraints: the base of the implanted anvil (boundary MN) is fixed, while the boundaries GH, HI, and IJ are prescribed to have no radial displacements with respect to the DAC axis, such that the symmetry axis in our model will not be shifted.

(3) Contact surfaces: the static Coulomb friction model is applied to simulate the contact between the anvils and the gasket, where boundaries UT, TS, PQ, and QR have a friction constant of 0.02 while boundaries SF and RJ have a friction constant of 0.2 (constants taken from Ref. \cite{Hsieh2019Imaging}).

To clearly portray the stress features in our DACs, the simulation results at the highest pressure point $p_6$ of DAC1 are summarized in \cref{fig4}(b) as examples, where the color maps visualize the spatial distributions of the loading stress $\sigma_{XX}$, the lateral stresses $\sigma_{YY}, \sigma_{ZZ}$, and one of the shear stresses $\sigma_{XY}$. Note that the distributions of $\sigma_{XX}$, $\sigma_{YY}$, and $\sigma_{ZZ}$ are symmetric in the two anvils but it is not the case for $\sigma_{XY}$. This may be due to the asymmetry of the gasket's pre-indentation and the fact that we compress the DAC from above. The shear stress $\sigma_{XY}$ is notably smaller than the diagonal stress components, in agreement with the simulation in Ref. \cite{Hsieh2019Imaging}. We notice, however, that in Ref. \cite{Hsieh2019Imaging}, they can reconstruct finite shear stresses from experimental data.

For each simulation, we average the simulated stress tensor components over the spatial region where the INVs are supposed to be in the DAC, i.e. 10 to 100 \textmu m from the DAC axis and 10 to 15 nm below the culet surface of the implanted anvil. These simulated results of INV crystal-frame stress tensor components are presented using lines in \cref{fig3}(c) and (d) for DAC1 and DAC2 respectively, which reveal the same gradual dominance of the loading stress over the lateral stresses as in our previous results derived from experimental data. Moreover, the simulation results substantiate our assumptions in \cref{sigmaINV2}: First, one of the shear stresses $\sigma_{XY, \mathrm{sim}}$ is negligible at all the pressure points under investigation; second, although the lateral stresses $\sigma_{YY, \mathrm{sim}}$ and $\sigma_{ZZ, \mathrm{sim}}$ exhibit different spatial dependencies (\cref{fig4}(b) as an example), they have very close averages over the INV region as shown by the overlapping dashed and dotted lines in \cref{fig3}(c) and (d). To conclude, the computational simulations uphold our speculation that the non-hydrostaticity of the stress environment inside the anvil culet mainly emanates from the dominant uniaxial stress along the DAC axis.

Results from \cref{sec:Compare,sec:tensor} elucidate that NDs perform better than INVs as hydrostatic pressure gauges. Given a hydrostatic pressure medium in a DAC, NDs at the medium interface efficiently receive the hydrostatic pressure, while INVs inside the anvil culet are heavily affected by the breaking of spatial symmetry due to the externally applied force. Moreover, our results substantiate that NDs have a longer working range for pressure detection compared with INVs. Throughout the hydrostatic pressure range of the pressure medium, NDs present tiny changes in $E^{\mathrm{net}}$ and contrast ratio of ODMR resonances. However, as the DAC pressure is increased, INVs show gradual suppression of one of the ODMR resonances due to non-hydrostaticity in the local stress environment, which hinders accurate pressure determination from the center of resonances. One way-out to extend the working range of INVs is to apply a magnetic field $\bm{B}$ of at least 50 Gauss along [100] of the diamond anvil, such that in the NV ground-state Hamiltonian, the magnetic field term $\gamma \norm{\bm{B}} \geq$ O(10$^2$) MHz is significantly greater than the $M_x^k$ and $M_y^k$ terms which are related to $E \leq$ O(10$^1$) MHz of INVs, where $\gamma=2.8$ MHz/Gauss is the gyromagnetic ratio for electrons. Then, the local non-hydrostaticity would bring negligible effects to the ODMR spectrum of INVs and we would have two ODMR resonances of similar contrasts. We can solve $D$ and hence $P$ from this spectrum using equations from the section of three-cone method in Supplementary Materials, with the known magnetic field projections on the NV orientations. Thus, a well-controlled magnetic field is necessary for INVs to work fine in the entire hydrostatic pressure range of the medium, while NDs do not require extra apparatus for robust pressure sensing.

\begin{figure}[t]
\includegraphics[width=8.6cm]{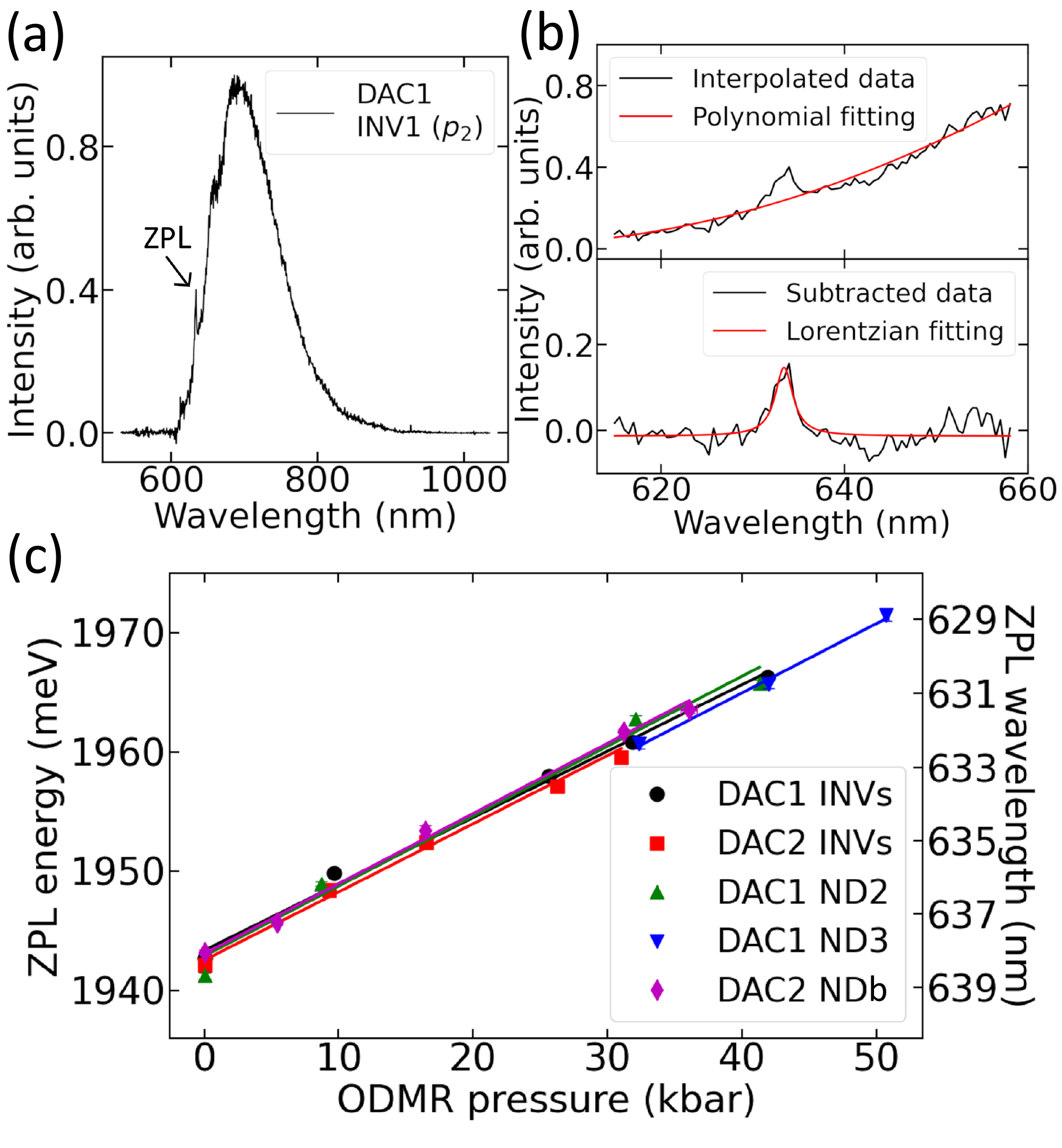}
\caption{(a) A sample NV PL spectrum obtained by the measurement method stated in \cref{sec:Exp}. In general, there is a narrow ZPL followed by broad phonon sidebands. (b) The ZPL fitting procedures in our data analysis, which concentrates on the portion of the PL spectrum from 615 to 658 nm. First, this piece of data is linearly interpolated and fitted with a quadratic polynomial ({\it Upper panel}). Then, the fitted polynomial is subtracted from the interpolated data and a Lorentzian peak fitting is performed ({\it Lower panel}). For both (a) and (b), the raw data from INV1 in DAC1 at $p_2$ is used as an example. (c) Linear fittings of ZPL energy versus ODMR-calibrated local pressure for the INVs and NDs in DAC1 and DAC2. The data points for INVs are averages from the six patches in the corresponding DAC. The five fittings reveal similar pressure dependencies of the ZPL, from 0.56 to 0.59 meV/kbar with errors on the order of 0.01 meV/kbar. Our experimental results agree well with the literature. }
\label{fig5}
\end{figure}

\section{\label{sec:ZPL} ZPL as an alternative pressure gauge}

Researchers have extensively studied the responses of NV ground-state spin sublevels to external perturbations, and developed the well-known ODMR spectroscopy for quantum information technologies. In fact, not only spin sublevels but also electronic orbitals of the NV center would be adjusted by perturbations. Here, we aim at quantifying the pressure-induced change in the energy spacing between electronic ground and excited states of the NV center, via measuring the pressure dependence of ZPL in the PL spectra of INVs and NDs. We expect the two types of NV sensors would concur on the dependence as long as their respective local pressures are calibrated by ODMR spectroscopy. 

By manipulating the NV spin state as described in \cref{sec:Exp}, we measure the NV PL spectra from the six (six) patches of INVs and two (one) of the NDs in DAC1 (DAC2) along the pressure sequence. A sample NV PL spectrum is shown in \cref{fig5}(a). In general, the PL spectrum of an NV ensemble consists of broad phonon sidebands trailing behind a narrow ZPL which undergoes a weak blue shift under pressure. Without delving into the complicated fitting of the entire PL spectrum, we consistently extract the ZPL position from each measured spectrum in the following steps: (i) focus on the data between 615 and 658 nm which fully captures the ZPL evolution in our experimental pressure range, (ii) linearly interpolate this portion of data and perform a quadratic polynomial fitting (see the upper panel in \cref{fig5}(b)), (iii) subtract the fitted polynomial from the interpolated data and perform a Lorentzian peak fitting (see the lower panel in \cref{fig5}(b)). Note that \cref{fig5}(b) depicts the ZPL fitting procedures using the raw data in \cref{fig5}(a) as an example.

With the ZPL positions distilled out, we linearly fit the curves of ZPL energy versus local pressure for our targeted sensors in DAC1 and DAC2 (see \cref{fig5}(c)), where the ZPL energy is converted from the extracted ZPL wavelength and the local pressure is determined by ODMR method. The data points for INVs are averages from the six patches in the corresponding DAC. The five fittings in \cref{fig5}(c) reveal similar slopes ranging from 0.56 to 0.59 meV/kbar, with errors on the order of 0.01 meV/kbar. This fulfills our earlier expectation that well-calibrated NV sensors should agree on the pressure dependence of the ZPL, and our results can match with the previously reported values \cite{Doherty2014Electronic, Doherty2014Temperature, Kobayashi1993High}. Moreover, it can be seen that our measured ZPL wavelengths are around 638 nm under ambient conditions, slightly off from the literature value of 637 nm. This may be due to some finite intrinsic stresses in the implanted diamond anvil. Note that each fitting in \cref{fig5}(c) only considers the pressure points where the corresponding sensor has sufficient ODMR contrasts to yield satisfactory NV PL spectra. Besides, the reduced pressure points are purposely excluded from the fittings in \cref{fig5}(c) for the same reason as before that relaxation of a DAC is not simply the reverse process of pressurization.

Our experimentally determined blue shift of the ZPL indicates a repulsion between NV electronic ground and excited states caused by pressure. In fact, it is of practical importance to confirm the slope of ZPL energy against pressure. By doing so, PL spectroscopy can be developed into an alternative to the ODMR method for pressure sensing with the NV center, and we can choose to utilize the spin or orbital degree of freedom in the NV energy structure for different experimental situations. PL spectroscopy is particularly useful if one does not want to introduce electrical feedthroughs into the DAC. An all-optical pressure sensing protocol is possible with PL spectroscopy, where one can obtain the NV PL spectrum by subtracting the spectrum measured near the NV sensor from the spectrum measured precisely at the location of the NV sensor, under an assumption of spatially uniform background PL signals. This assumption is valid if no components in the DAC other than NV sensors would emit red fluorescence under green laser excitation. With ODMR and PL spectroscopies, the NV center can be a resilient pressure sensor that caters to different experimental conditions.

\begin{figure}[t]
\includegraphics[width=8.6cm]{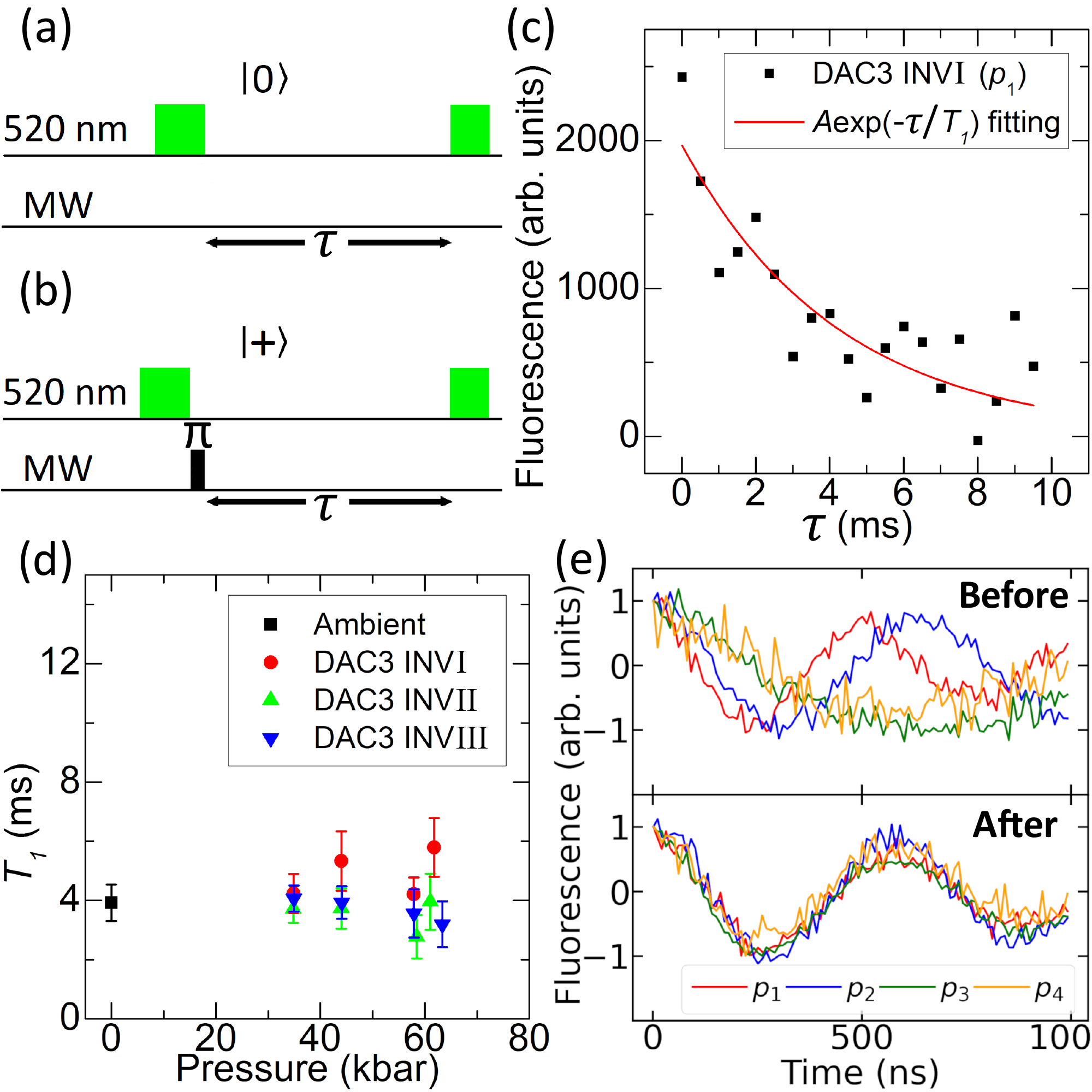}
\caption{(a, b) Pulse sequences for measuring the relaxation curves of the bright $\ket{0}$ state and the dark $\ket{+}$ state respectively. The 520-nm laser is for initialization to the $\ket{0}$ state and readout of the final state, while the MW $\pi$ pulse is to flip $\ket{0}$ to $\ket{+}$. Furthermore, $\tau$ is the free precession time to be varied, and the $\pi$ pulse is half the Rabi period of driving the right-hand ODMR resonance $f_{+}$. (c) Exponential decay fitting of the net relaxation curve obtained by subtracting the $\ket{+}$ curve from the $\ket{0}$ curve. The fitted decay time is taken as the experimental spin-lattice relaxation time $T_1$. The data from INV\RNum{1} in DAC3 at $p_1$ is used as an example here. (d) $T_1$ time versus ODMR-calibrated local pressure for three patches of INVs in DAC3, with the ambient-condition data from a patch of INVs close to the six tracked patches as a reference. No significant changes in $T_1$ time are observed. (e) The Rabi oscillations at different pressure points before and after tuning the MW power fed into the DAC. ({\it{Upper panel}}) If the input MW power is kept the same in the experiment, the Rabi period changes with the DAC pressure, which is expected due to the varying MW transmission efficiency through the Omega-shaped antenna when the DAC is pressurized. ({\it{Lower panel}}) After appropriately tuning the input MW power, the Rabi period can be fixed at 582 ns such that the INVs receive the same MW power at all pressure points. Here, the data from INV\RNum{1} in DAC3 is used as an example, and the Rabi oscillation amplitudes are normalized for easier comparison of periods.}
\label{fig6}
\end{figure}

\section{\label{sec:T1} Pulsed measurements with a hydrostatic pressure medium}

Pulsed measurements are key to enhancing the sensitivity and realizing complex sensing schemes \cite{Barry2020Sensitivity}. In real life, spin decoherence creates difficulties in implementing pulse sequences. The decoherence occurs via two channels: (i) the relaxation in $z$-direction of the Bloch sphere due to electron-phonon coupling between NV centers and the lattice; (ii) the dephasing in $x$-$y$ plane of the Bloch sphere due to spin-spin interactions. These two decoherence channels are characterized by the $T_1$ and $T_2$ times, respectively. In recent years, some promising pulsed sensing protocols have been demonstrated in either ambient or pressurized conditions \cite{Hsieh2019Imaging, McLaughlin2021Strong, Monge2022Spin, Tetienne2013Spin, Schmid-Lorch2015Relaxometry, Wang2021Ac}. Nonetheless, little attention has been paid to the hydrostaticity of the pressure medium and the characterization of NV decoherence times using a hydrostatic medium. These concerns are important for high-fidelity NV sensing and NV-based quantum computing to be robustly performed in extreme conditions. To address these concerns, we examine the INV $T_1$ time over the course of pressurizing our DACs, where we check the hydrostaticity of the pressure medium with great caution.

Knowing the results in \cref{sec:Compare,sec:tensor}, one may wonder if the local stress anisotropy in the diamond anvil culet would induce peculiar crystal deformations and modify the system’s electron-phonon coupling which in turn affects the $T_{1}$ time of the INVs. Thus, we would focus on the INVs rather than the NDs in this subsection. Refer to \cref{sec:Theory}, if the shear stresses are negligible (suggested by the simulations in \cref{sec:tensor}), all the four NV orientations would have the same eigenfrequencies, $f_0=0$ and $f_{\pm}$, and the same eigenstates, $\ket{m_s=0}$ and $\ket{\pm}$, where $\ket{\pm}$ are superpositions of $\ket{m_s=+1}$ and $\ket{m_s=-1}$. In this case, we would like to study the $T_1$ time of the two-level system spanned by $\ket{0}$ and $\ket{+}$ with transition frequency $f_{+}$, which can be obtained by fitting the right-hand ODMR peak at zero magnetic field. With the pulse sequences depicted in \cref{fig6}(a) and (b), we can measure the relaxation curves of the bright $\ket{0}$ state and the dark $\ket{+}$ state respectively. Note that the $\pi$ pulse in \cref{fig6}(b) is half the Rabi period of driving the resonance $f_{+}$. To extract the $T_1$ time, we subtract the relaxation curve of $\ket{+}$ from that of $\ket{0}$, and fit the resulting curve with an exponential decay $A \exp(-\tau/T_1)$ (see \cref{fig6}(c) as an example), where the amplitude $A$ is limited by the Rabi contrast. Here, we focus on the right-hand ODMR resonance because the left-hand ODMR contrast of INVs is suppressed when the DAC pressure is increased as shown in \cref{fig1}(d).

We perform the $T_1$ measurement using a new DAC (named as DAC3) which has the same cell configurations as DAC2 but with 99.5\% glycerol as the pressure medium. Compared with 4:1 methanol:ethanol mixture, glycerol is a more common medium since it is not a strong solvent, and it is chemically inert. The ascending pressure sequence for DAC3 is from the ambient pressure $p_0$ to $p_4$ without any reduced pressure points. Through inspecting the SD of pressure among NDs, we find that our prepared glycerol in DAC3 has a critical pressure $P_{c}$ at around 80 kbar (see Supplementary Materials), which is in good agreement with \cite{Yip2019Measuring}. Our medium is therefore perfectly hydrostatic before $p_4$ and quasi-hydrostatic at $p_4$. By tracking two NDs and six patches of INVs in DAC3, we can reproduce the results in \cref{fig1,fig2} (see Supplementary Materials), proving that our previous claims are independent of the choice of pressure medium as long as it is in the hydrostatic regime. For DAC3, we will number the tracked sensors using Roman numerals, e.g. ND\RNum{1}, INV\RNum{1}.

We have monitored the $T_1$ time for three patches of INVs in DAC3 from $p_1$ to $p_4$, with the ambient-condition data from a patch of INVs close to the six tracked patches as a reference (INVs in the implanted region exhibit consistent ODMR properties from previous experience with DAC1 and DAC2). The measurement results are shown in \cref{fig6}(d), where no significant changes in the $T_1$ times are observed. Note that the efficiency of MW transmission through the Omega-shaped antenna is inevitably changed when we increase the DAC pressure. To ensure a constant MW power received by the INVs at different pressure points for fair comparison of the $T_1$ times in \cref{fig6}(d), we have fixed the Rabi period to be 582 ns for all patches of INVs at all pressure points by tuning the MW power fed into the DAC. We assume negligible detuning in the Rabi oscillations here. As examples, \cref{fig6}(e) shows the Rabi oscillations of a patch of INVs in DAC3 at different pressure points before and after tuning the input MW power.

Our measurement results indicate that possible modifications to the electron-phonon coupling by the local stress anisotropy in the anvil culet are tiny and within our experimental errors, given the medium is in a good hydrostatic condition. This demonstrates the stability of NV properties under extreme conditions and once again proves the robustness of NV sensing. A natural extension of our work is to monitor the $T_2$ time of INVs under pressure by implementing the Hahn-echo pulse sequence, but to do so, a well-controlled magnetic field is a requisite for aligning a magnetic field along one of the NV orientations. Our preliminary results (data not shown) under zero magnetic field in DAC3 seem to indicate no observable changes in the $T_{2}$ time up to 80 kbar.

\section{\label{sec:nanopillar} Proposal of diamond nanopillars in a DAC}

\begin{figure}[t]
\includegraphics[width=8.6cm]{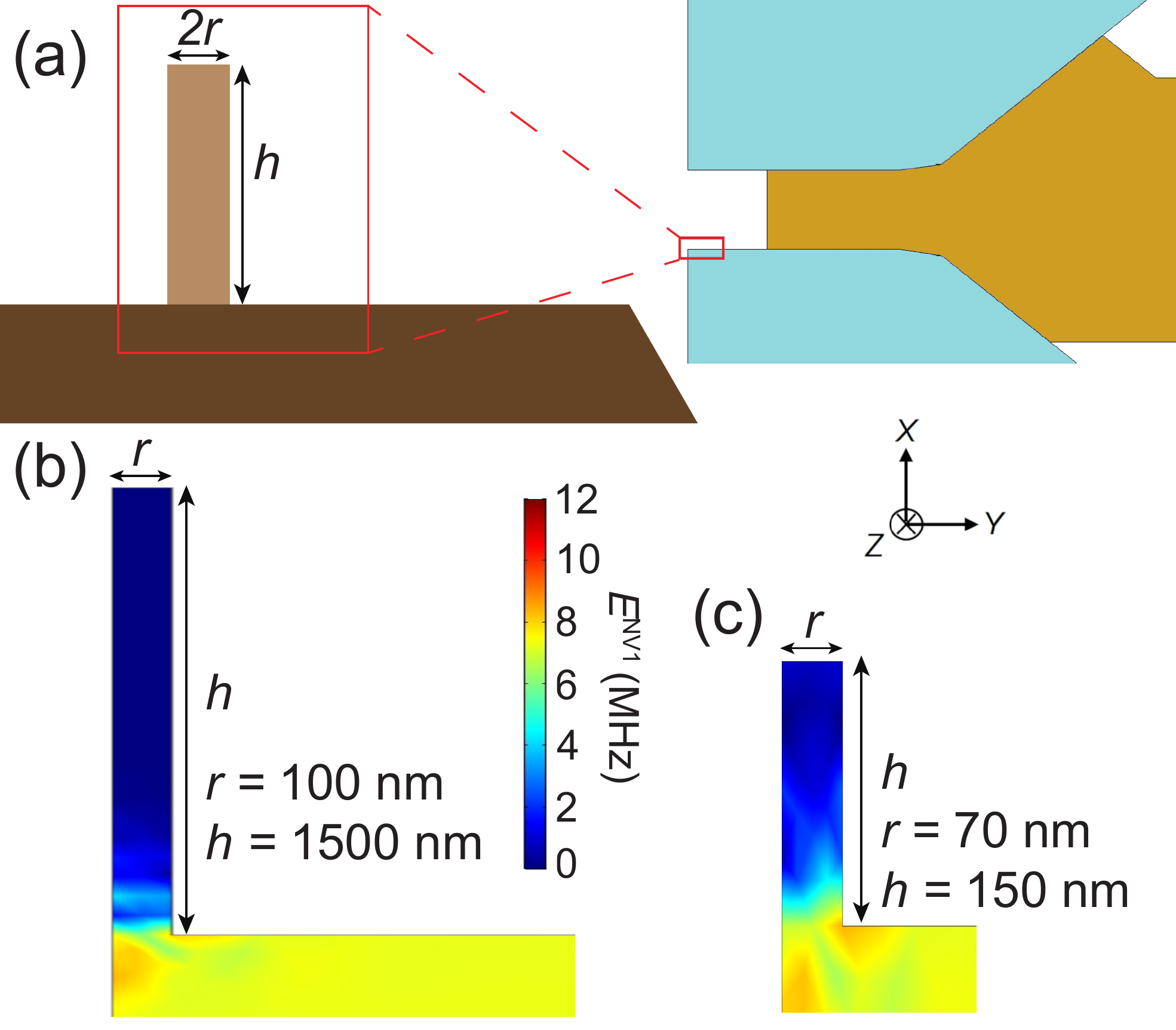}
\caption{(a) The modified 2D axisymmetric DAC model with a diamond nanopillar at the center of the implanted anvil culet. The zoom-in is a schematic diagram showing a nanopillar, where $r$ and $h$ are the radius and the height, respectively. (b, c) Spatial maps of $E^{\mathrm{NV1}}$ derived from the simulations using the nanopillar geometry of $r,h=100,1500$ nm and the geometry of $r,h=70,150$ nm, respectively. The two maps share the same color scale. $E^{\mathrm{NV1}}$ is found to be negligible within the nanopillar but much larger inside the bulk of the anvil.}
\label{fig7}
\end{figure}

In addition to using NDs and applying a bias magnetic field, we propose here the third method to mitigate the adverse effects of the non-hydrostaticity in the pressurized environment and to extend the working pressure of NV centers as quantum sensors. The fabrication and characterization of diamond nanopillars have been discussed in the literature \cite{Momenzadeh2015Nanoengineer, Widmann2015Fabrication, Blake2020Plastic, Zou2008Fabrication}, and NV sensing using nanopillars has been demonstrated under ambient conditions \cite{Momenzadeh2015Nanoengineer}. We therefore suggest the integration of nanopillars into a DAC so that the INVs embedded in the nanopillars can perform reliable quantum sensing in a hydrostatic stress environment. We conduct finite-element simulations to support our proposal, where we modify the 2D axisymmetric model in \cref{fig4}(a) by adding one nanopillar to the center of the implanted anvil culet (the bottom anvil in \cref{fig4}(a)). For the 2D model, we use the dimensions of DAC1 and try two different nanopillar geometries. One geometry has a radius of 100 nm and a height of 1500 nm, as inspired by the nanopillars in Ref.\cite{Momenzadeh2015Nanoengineer}. The other geometry has a radius of 70 nm and a height of 150 nm, which is similar to the size of our 140-nm NDs. We simulate the stress distribution under the boundary conditions stated in \cref{sec:tensor}, where the boundary loads follow the data at the highest pressure point $p_6$ of DAC1. In \cref{fig7}(b) and (c), we map the spatial distribution of $E^{\mathrm{NV1}}$ in the implanted anvil for the two geometries, revealing that the indicator of non-hydrostaticity, $E$, is close to zero within the nanopillar but it is much larger inside the bulk of the anvil. This hints at a good hydrostatic condition in the nanopillar, such that the mixing of NV ground-state spin sublevels can be minimized and the INVs inside the nanopillar can perform robust quantum sensing based on ODMR spectroscopy. 

\section{\label{sec:summary}Discussion and Summary}
Integrating different NV sensor types has some useful sensing applications. For instance, one may integrate INVs and NDs to study a liquid-solid phase boundary, where the liquid and solid properties can be sensed by immersed NDs and shallow INVs respectively. Understanding liquid-solid interfaces at a microscopic scale is a prevailing challenge in quantum chemistry \cite{Zaera2012Probing}, and NV sensing may provide new opportunities to the field.

In summary, this work has revealed a noticeable difference in the local stress environments encountered by INVs and NDs in the same DAC. Note that NV is just a platform for reconstructing the stress components, and our results should be generalized to the stress discrepancy between different parts in a high-pressure device: more hydrostatic at the pressure medium interface and more anisotropic inside the force-transmitting solid elements, given a hydrostatic pressure medium below $P_\mathrm{c}$. Moreover, our experiments and simulations demonstrate the sensitivity of NV centers to different stress profiles. Although INVs can be a versatile non-invasive tool in diamond-based pressure devices, NDs appear to be a better option for gauging hydrostatic pressure and have a longer working range for pressure detection with zero magnetic field. In fact, any type of NV sensor can be a legitimate pressure gauge as long as it is well-calibrated, and our work is exactly dedicated to characterizing the behaviors of different NV sensors in a confined pressure device. We want to emphasize that the choice of NV sensors heavily depends on the experimental purpose so that their unique advantages can be fully utilized. Furthermore, this work provides insights on different aspects of the NV energy structure. We confirm a pressure-induced repulsion between NV electronic ground and excited states by measuring the blue shift of ZPL in the NV PL spectrum. We also show that the electron-phonon coupling in the NV system would not be significantly modified by local stress anisotropy, as seen from the measured stability of the INV $T_1$ time. With a deeper understanding of the pressure-tuned NV system, more different sensing applications of the NV center are expected in the future.

Our work also addresses the tolerance to non-hydrostaticity when NV centers are applied as versatile sensors in pressurized environments, which is a key question from the NV community. Non-hydrostaticity hinders NV sensing in the following ways: (i) the ground-state spin sublevels are mixed in the energy eigenstates under a non-hydrostatic stress field (see \cref{Hk}), lowering the efficiency of ODMR spectroscopy; (ii) magnetic field sensing using NV centers would be inaccurate if the $E$ term in \cref{E} is comparable to the magnetic field. Our empirical results show that when the indicator of non-hydrostaticity, $E$, reaches O(10$^1$) MHz, one of the resonances in the ODMR spectrum of NV centers at zero bias field is heavily suppressed, and the sensing accuracy is thus decreased. This would impose restrictions on the maximum working pressure of NV sensing. There are three solutions when we encounter this threshold of $E\sim$ O(10$^1$) MHz. A straightforward solution would be to use NDs in a pressure medium with a sufficiently high $P_\mathrm{c}$ to ensure a hydrostatic environment around the NV centers. Another solution would be to fabricate a diamond anvil with some nanopillars on the anvil culet. The NV centers encompassed in the nanopillars would perceive a more hydrostatic environment similar to NDs drop-casted on the culet, as demonstrated in the finite-element simulations in Supplementary Materials. Besides, the crystal orientation of the nanopillers is definite in the laboratory frame, which is an advantage over NDs. The third solution would be to apply a bias magnetic field of $\gamma \norm{\bm{B}} \gg E$ such that the non-hydrostaticity is not the dominant term in the Hamiltonian, which has been demonstrated in the literature \cite{Hsieh2019Imaging, Lesik2019Magnetic}.

\begin{acknowledgments}
We thank P. T. Fong for the fruitful discussion. K.O.H acknowledges financial support from the Hong Kong PhD Fellowship Scheme. S.K.G. acknowledges financial support from Hong Kong RGC (GRF/14300418, GRF/14301020, and A-CUHK402/19). S.Y. acknowledges financial support from Hong Kong RGC (GRF/14304419).

Kin On Ho, Man Yin Leung, and Wenyan Wang contributed equally to this work.
\end{acknowledgments}

\bibliography{references}

\end{document}